**Piezomagnetism-driven magnetoelectric coupling in altermagnetic multiferroic $K_3Cr_2F_7$**

Ying Zhou[1], Hui-Min Zhang[1], Cheng-Ao Ji[1], Hongjun Xiang[2,3], Shuai Dong[1], James M. Rondinelli[4,*] and Xue-Zeng Lu[1,*]

[1]Key Laboratory of Quantum Materials and Devices of Ministry of Education, School of Physics, Southeast University, Nanjing 211189, China.

[2]Key Laboratory of Computational Physical Sciences (Ministry of Education), Institute of Computational Physical Sciences, and Department of Physics, Fudan University, Shanghai 200433, China.

[3]Shanghai Qi Zhi Institute, Shanghai 200030, China.

[4]Department of Materials Science and Engineering, Northwestern University, Evanston, Illinois 60208, USA.

*jrondinelli@northwestern.edu
*xuezenglu@seu.edu.cn

**Abstract**

Ferroelectric control of altermagnetism in momentum space has been studied widely, while the control of magnetism in real space of altermagnets are still rare. We present a design rule to identify multiferroicity in $n$=2 Ruddlesden-Popper halides. Our results show that a Jahn-Teller distortion can cooperate with oxygen octahedral rotations to break inversion symmetry, which we demonstrate in $K_3Cr_2F_7$ and cation-ordered $KAg_2Cu_2Cl_7$, and leads to a ferrielectric-to-ferroelectric phase transition in $K_3Cr_2F_7$. Altermagnetic spin order in the ferrielectric phase of $K_3Cr_2F_7$ transforms into a conventional antiferromagnetic order in the ferroelectric phase, at which strain/pressure engineered sizable changes of weak ferromagnetism can occur. Our study is not only conducive to realize strong magnetoelectric coupling in multiferroics, but also reveals more functionalities in altermagnetic materials.

**Introduction**

Altermagnetic spin order was originally proposed in collinear antiferromagnetic (AFM) materials and extended to noncollinear AFM materials recently(*1–8*). The importance of these materials is that they have electronic bands splitting, separating the spin-up and spin-down bands in momentum space, without spin-orbital coupling (SOC). Computational studies have identified many centrosymmetric and noncentrosymmetric altermagnetic materials (*9*) and experimental measurements of the bands splitting have been performed in some of them by using spin-resolved and angle-resolved photoemission spectroscopy techniques(*3, 8*).

Furthermore, recent experiments discovered a new type of spin texture that lifts time-reversal symmetry(*8*), i.e., the quadratic spin texture, distinct from the time-reversal symmetry conserved Rashba and Dresselhaus spin textures(*10, 11*). Two magnetic domains mapped by time-reversal symmetry in altermagnets (*6*) can also lead to opposite crystal Hall effects and piezomagnetic effects(*12*). Very recently, the ferroelectric control of altermagnetism in the momentum space has been reported and attracted much attention(*13–15*).

In magnetoelectric multiferroics, the coupling required for realizing electric field control of magnetism has been extensively studied in many Type-I and Type-II multiferroics(*16–18*). However, how the unique properties of altermagnetic materials might facilitate such electric field control remains largely unexplored.. Here we present first-principles calculations and group theory analysis of Ruddlesden-Popper (RP) materials, which are well known for hosting improper ferroelectricity induced by octahedral rotations [i.e., hybrid improper ferroelectricity (HIF)]; thus, leading to possible strong magnetoelectric (ME) coupling not yet reported experimentally(*19, 20*). Our results demonstrate that incorporating Jahn-Teller (JT) active magnetic ions into RP compounds generates an alternative polar phase having a group-subgroup relationship to the previously observed polar $Cmc2_1$ phase. Through our group theory analysis and electronic structure calculations, we show the new polar phase hosts ferrielectricity and altermagnetism. The coexisting ferrielectric and altermagnetic state can be transformed into a phase with ferroelectric and conventional AFM order (i.e., $Cmc2_1$) with a low energy barrier, which leads to strong ME coupling between the electric polarization and weak ferromagnetism (wFM) tuned by strain or pressure. Finally, we propose a design rule to facilitate the finding of such ferrielectric phase in RP materials.

**Computational details**

Our total energy calculations are based on density functional theory (DFT) within the generalized gradient approximation (GGA) utilizing the revised Perdew-Becke-Erzenhof functional for solids (PBEsol) (*21*) implemented in the Vienna Ab Initio Simulation Package (VASP)(*22, 23*).

We use a 550-eV plane wave cutoff energy for all calculations and the projector augmented wave (PAW) method, and a Γ-centered 5× 5 × 2 $k$-point mesh for the Brillouin zone integration. We use Gaussian smearing (0.10 eV width) for the Brillouin-zone integrations and a sufficiently dense $k$-point mesh. The pseudopotentials for the $A$- and $B$-site ions are chosen to treat the $s$ or $p$ core electrons as the valence electrons if available. The DFT plus Hubbard $U$ method is used(*24*), and the $U$ value is tested from 3 to 5 eV (**Table S1** in the **Supplementary Material**(*25*)) for the $K_3Cr_2F_7$. The Hubbard $U$ and the exchange parameter $J$ set to 5 eV and 1 eV for Cr, respectively, which is tested for describing $K_3Cr_2F_7$. The $U$ value is tested from 4 to 6 eV (**Table S4** and **S5**(*25*)) for $KAg_2Cu_2Cl_7$ and the $K_3Cu_2Cl_7$. Spin-orbital coupling (SOC) is included for the magnetic anisotropy studies. The ferroelectric polarization was calculated using the Berry phase method(*26*).

To compute the symmetric spin exchange parameters, we use the four-state mapping method(*27, 28*). We calculate the effective symmetric spin exchange parameters, which are obtained by setting $|S_i|$=1, namely, $J_{ij} = J_{ij}^{eff} S_i S_j$ for a spin dimer $ij$.

Our parallel tempering Monte Carlo (PTMC) simulations are based on an exchange MC method(*29, 30*), which can simulate the classical Heisenberg spin system with a Hamiltonian $E = E_0 + \sum_{i,j} J_{ij} S_i \cdot S_j$, where $J_{ij}$ is the symmetric spin exchange parameter. To obtain the plot of the specific heat ($C$) versus temperature ($T$), we calculate the specific heat $C \sim (\langle E^2 \rangle - \langle E \rangle^2)/T^2$ after the system reaches equilibrium at a given temperature $T$ in the simulation. Then, we obtain the critical temperature by locating the maximum in the $C(T)$ plot. In our PTMC simulations of the effective Hamiltonian, a 6×6×3 supercell of the 48-atom unit cell is adopted for $K_3Cr_2F_7$, which we confirmed is converged.

For our phonon calculations, we use the finite displacements method (*31*). Phonon dispersion calculations at the finite temperatures were performed using the Temperature Dependent Effective Potential (TDEP) method(*32, 33*). The symmetry-adapted mode distortion vectors are obtained by performing a mode decomposition using the ISODISTORT software(*34*).

The tolerance factor is obtained by using $\tau_i = \frac{r_A + r_X}{\sqrt{2}(r_B + r_X)}$, where $r_A$, $r_B$ and $r_X$ are the ionic radii of the $A$, $A'$, $B$ cations and anion, respectively. The averaged $r_{A-X} = r_A + r_X$ is computed by considering the nearest-neighboring oxygen atoms about the $A$ atom for all $A$ and $A'$ atoms. The averaged $r_{B-X}$ is computed in the same way as $r_{A-X}$. Here, we consider 12-coordination for $A$, 9-coordination for $A'$ and 6-coordination for $B$.

**Results**

The first all-inorganic multiferroic $n=2$ RP halide (i.e., (Rb, K)$_3$Mn$_2$Cl$_7$) was recently discovered and it exhibits both HIF and G-type AFM spin orders at ~ 64 K(*35*). With this as our prototype material, we carry out cation and anion substitutions and obtain more $n=2$ RP halides. Taking K$_3$Cr$_2$F$_7$ as an example, our results indicate that the phases with space groups *Pnma* (No. 62) and *Cmc*2$_1$ (No. 36) exhibit relatively low energies [**Fig.1(a)**]. Phonon modes decomposition reveals that the *Cmc*2$_1$ structure is composed of the in-phase octahedral rotational modes $X_2^+$ [irreducible representation (irreps)] with an order parameter direction (OPD) (a, 0), out-of-phase octahedral tilt mode $X_3^-$ with OPD (a, 0), and in-plane ferroelectric (FE) mode $\Gamma_5^-$ with OPD (a, a). The *Pnma* structure consists of the modes $X_2^+$ with OPD (a, 0), $X_3^-$ with OPD (0, a), and antiferroelectric (AFE) mode $M_5^-$ with amplitude (a, 0) [**Fig. 1(b)**]. All decomposed modes are obtained with respect to the high-symmetry *I*4/*mmm* phase. It is further found that there is an imaginary-frequency mode with irrep Y$_3$ in the phonon spectrum of the *Cmc*2$_1$ phase [**Fig. 2(a)**]. Upon freezing this Y$_3$ distortion mode in the *Cmc*2$_1$ phase, we obtained the *Pmn*2$_1$ phase, which is dynamically stable [**Fig. 2(b)**]. Meanwhile, our computed phonon spectra at 50 K, obtained by using the TDEP method, demonstrate that both phases can be dynamically stable (**Fig. S1**(*25*)). Therefore, our results show that the pure RP compound K$_3$Cr$_2$F$_7$ with the JT active Cr$^{2+}$ ion can have a polar ground state phase of *Pmn*2$_1$ symmetry.

To further elucidate the origin of the appearance of the *Pmn*2$_1$ polar phase, we decompose the structure with respect to the high-symmetry *I*4/*mmm* phase. In *Pmn*2$_1$ phase, there are five main distortion modes: $\Gamma_5^-$, $M_5^-$, $X_2^+$, $X_3^-$, and a JT distortion mode ($X_1^+$), as illustrated in **Fig. 1(b)**. The *Pmn*2$_1$ structure contains an additional distortion mode absent in either the *Cmc*2$_1$ or *Pnma* phases; namely, the $X_1^+$ mode that corresponds to the JT distortion of the octahedra. Furthermore, the $X_3^-$ mode with a OPD (a, b) in the *Pmn*2$_1$ phase also differs from those in both the *Cmc*2$_1$ and *Pnma* phases, which plays an important role in stabilizing the ferrielectric *Pmn*2$_1$ phase with both FE and AFE modes. As can be seen in **Fig. 3(a)**, both FE $\Gamma_5^-$ and AFE $M_5^-$ are not energy lowering when adding them individually into the *I*4/*mmm* phase. In contrast, the $X_2^+$ and $X_3^-$ modes exhibit significant energy-lowering. This indicates that both *Cmc*2$_1$ and *Pmn*2$_1$ can be hybrid improper ferroelectrics. Among the five main phonon modes in the *Pmn*2$_1$ phase, there are four trilinear coupling interactions [**Fig. 3(b)**]: $Q_{\Gamma_5^-}Q_{X_2^+}Q_{X_3^-}$, $Q_{M_5^-}Q_{X_2^+}Q_{X_3^-}$, $Q_{\Gamma_5^-}Q_{X_1^+}Q_{X_3^-}$ and $Q_{M_5^-}Q_{X_1^+}Q_{X_3^-}$. We also compute the energy gain from the four trilinear coupling interactions in the *Pmn*2$_1$ phase, as shown in **Fig. 3(b)**. The trilinear coupling interactions in the *Pmn*2$_1$ phases also largely contribute to the energy lowering, which demonstrates the importance of the trilinear coupling interactions in stabilizing the *Pmn*2$_1$ phase.

We also investigate the other complex fluorides comprising $Mn^{2+}$ ($d^5$, JT-inactive) and $Cu^{2+}$ ($d^9$, JT-active), such as $K_3Mn_2F_7$ and $K_3Cu_2Cl_7$. They exhibit the nonpolar *Cmcm* and *Pbcn* phases, respectively [**Table 1**], despite having tolerance factors close to $K_3Cr_2F_7$. By further lowering the tolerance factor with the A-site cation ordering, we can obtain the polar *Pmn*2$_1$ phase in $KAg_2Cu_2Cl_7$, but not in the cation ordered manganese halides. It seems that both tolerance factor and JT activity are important to enable the ground state *Pmn*2$_1$ phase.

From the previous studies, we can obtain the following trends in the changes of the octahedral rotations and tilts with decreasing tolerance factor: (1) $X_3^-$ mode in *P4$_2$/mnm* with OPD (a,a) first appears; (2) $X_3^-$ mode in *Cmcm* with the OPD (a,0) or $X_1^-$ mode (out-of-phase octahedral rotation) in *Ccca* phase appears; (3) both $X_3^-$ (a,0) and $X_2^+/X_1^-$ modes appear [*Cmc*2$_1$ ($X_3^-$ and $X_2^+$) or *Pbcn* ($X_3^-$ and $X_1^-$)]; and (4) $X_3^-$ (a,0), $X_2^+$ and $X_1^-$ modes can occur leading to the *Pc* phase. This sequence was observed in halide(*36*) and oxide RP materials(*37–44*). The ground state phases of the materials in **Table 1** also follow the sequence described above except $K_3Cr_2F_7$ and $K_3Cu_2Cl_7$. It can be found that in $K_3Cr_2F_7$, there can also be octahedral rotations $X_2^+$ and $X_3^-$ (a, b) with the large tolerance factor (~0.97) at which *P4$_2$/mnm*, *Cmcm* or *Ccca* are usually stable in the RP materials with JT-inactive ions, leading to the *Pmn*2$_1$ ground state phase. In $K_3Cu_2Cl_7$, there can be octahedral rotations $X_1^-$ and $X_3^-$ (a, 0) with the large tolerance factor, leading to the *Pbcn* phase. This further indicates that the prediction on the polar ground state phase based on the tolerance factor will be so different by including the JT active transition-metal ion.

To investigate the appearance of the *Pmn*2$_1$ polar phase upon changing the tolerance factor, we compute the energies of *Pnma*, *Cmc*2$_1$ and *Pmn*2$_1$ phases, on the basis of a Landau model in $K_3Cr_2F_7$, expressed in *Pnma* and *Cmc*2$_1$ as:

$$E=E_0+\alpha_i Q_i^2 + \beta_i Q_i^4 + \alpha_j Q_j^2 + \beta_j Q_j^4 + \alpha_k Q_k^2 + \beta_k Q_k^4 + h_{i0} Q_i Q_j Q_k + \text{higher orders} \tag{1}$$

where $E_0$ is the energy of the *I4/mmm* structure, $Q_i$, $Q_j$ and $Q_k$ represent the order parameters in unit of Å of $i=\Gamma_5^-$ ($M_5^-$), $j=X_2^+$ and $k=X_3^-$ distortions in *Cmc*2$_1$ (*Pnma*), respectively, and $\alpha$, $\beta$, $h$ and $\gamma$ are the coefficients. In the *Pmn*2$_1$ phase, the Landau model can be written as:

$$E=E_0+\alpha_i Q_i^2 + \beta_i Q_i^4 + \alpha_j Q_j^2 + \beta_j Q_j^4 + \alpha_k Q_k^2 + \beta_k Q_k^4 + \alpha_q Q_q^2 + \beta_q Q_q^4 + \alpha_p Q_p^2 + \beta_p Q_p^4 + h_{i0} Q_i Q_j Q_k + h_{i1} Q_q Q_j Q_k + h_{i2} Q_i Q_p Q_k + h_{i3} Q_q Q_p Q_k + \\ +higher\ orders \tag{2}$$

where $Q_i$, $Q_j$, $Q_k$, $Q_q$ and $Q_p$ represent the order parameters in unit of Å of $i=\Gamma_5^-$, $j=X_2^+$, $k=X_3^-$, $q=M_5^-$ and $p=X_1^+$ distortions in $Pmn2_1$. To more accurately obtain the energies of the $Pnma$, $Cmc2_1$ and $Pmn2_1$ phases appearing in **Fig. 1(a)**, we minimize the Landau expansion by directly carrying out the DFT calculations rather than fitting the coefficients in the expansion [**Fig. 3(c)**]. Here the selectively normalized $Q$ modes considered in equations (1) and (2) are constrained, that is, the magnitude of the order parameter is equal to unity and the order parameter direction is fixed, however, the internal atomic positions are allowed to be optimized according to the directions of the $Q$ modes considered (that is, the magnitude of $Q$ is relaxed in the optimization) and the lattice constants are also optimized (*45*).

To account for the effects of tolerance factor that is closely related to the octahedral rotations in the perovskite-related materials, we fix the $X_2^+$ mode and relax the other modes in the three phases, because all the three phases have the same kind of $X_2^+$ (a,0) mode. As shown in **Fig. 3(c)**, the small $X_2^+$ mode will favor the competition of the $Pmn2_1$ phase with the $Pnma$ and $Cmc2_1$ structures. This explains that the $Pmn2_1$ phase, with the smallest $X_2^+$ mode among the three phases, is more stable than the others. To further demonstrate the reliability of the Landau model analysis, we model cation-ordered $KAg_2Cr_2F_7$ with a small tolerance factor of 0.961 and find the ground state phase becomes $Pnma$ (**Table 1** and **Table S6**(*25*)), because the energy differences between the $Pmn2_1$ phase and the other two phases become larger at the smaller tolerance factor (i.e., larger octahedral rotational modes). The further stabilization of the $Pnma$ phase may arise from the fact that the $Pnma$ phase has a smaller $X_2^+$ mode (1.66 Å) than that (1.71 Å) of the $Cmc2_1$ phase when fully relaxing the structures in the DFT calculations in $KAg_2Cr_2F_7$. Data in **Fig. 3(c)** supports this, and the energies of the three phases increase with increasing the amplitude of the $X_2^+$ mode and the $Cmc2_1$ and $Pnma$ are closely competing in energy.

Although we identified a route to stabilize the $Pmn2_1$ phase in $K_3Cr_2F_7$, i.e., increasing the tolerance factor favors its formation, our investigations of $K_3Cu_2Cl_7$ and $KAg_3Cu_2Cl_7$ reveal that, in these cases, the $Pmn2_1$ phase becomes stable only when the tolerance factor is reduced, allowing it to compete effectively with the $Pnma$, $Pbcn$, $Cmc2_1$, and $Pc$ phases. While we cannot draw a unified conclusion for predicting the ground-state structures of different $n=2$ RP materials with including the JT active ion, the finding of a $Pmn2_1$ phase in $n=2$ RP compounds largely relies more on the octahedral rotations. Thus, cation/anion substitution should be an effective route for tuning the tolerance factor (equivalently, tuning of the octahedral rotations) of the structure and can help find additional polar $Pmn2_1$ phases with other elements, such as in the RP oxides (*20, 46–49*). Last, the origin of the JT-induced polar phase found here is different from the previous studies in which the JT effects arise in cooperation with other microscopic driving forces for sustaining an electric polarization, such as layered *A*-site cation/organic ligand ordering and extremely large tensile strain (>4%)(*50–53*). These extra conditions are challenging to realize in experiments.

Next, we investigate the magnetic properties of *Pmn*2$_1$ in K$_3$Cr$_2$F$_7$. Since JT active Cr$^{2+}$ leads to a C-type orbital ordering with alternatively occupied perpendicular $d_{z^2}$ orbitals [**Fig. 4(a)**], the magnetic structure compatible with such C-type orbital ordering is A-type spin order (**Table S2 and Fig. S2** (*25*)). By including SOC, the magnetic anisotropy (MA) can be obtained. The unoccupied $d_{x^2-y^2}$ orbital will interact with the occupied $d_{yz}$ and $d_{xz}$ orbitals (**Fig. 4(b) Fig. S3** (*25*)) and this will result in an easy-plane single-ion anisotropy in the local coordinate(*54*) (**Fig. S4(a)** (*25*)). Since all octahedra have the local *z* axis close to the out-of-plane direction of the crystal structure, the magnetic structure has in-plane MA. Along with the above analysis, our DFT calculations further demonstrate that the ground state magnetic structure has *Pmn*2$_1$ magnetic group symmetry with the MA along the polarization direction (i.e., *x* direction), in which there is (A, A, C) spin ordering in each double-perovskite layer that leads to zero wFM (**Fig. 5(a), Fig. S7 and Table S8** (*25*)). By analyzing the symmetry of the *Pmn*2$_1$ magnetic structure, we find the spin up and spin down sublattices are connected by a spin space symmetry $[C_2||\{m_{001}|t\}]$. There are no inversion and translational symmetries between all the antiparallel spins [**Fig. 4(c)**]. Thus, this phase fulfills the requirements for exhibiting altermagnetism, which we confirm in our band structure calculations without SOC. It is evident that along the $m_{001}$-connected Brillouin zone paths Y1→Γ [from (0.0, 0.3, 0.5) to (0.0, 0.0, 0.0)] and Y1-→Γ [from (0.0, 0.3, -0.5) to (0, 0, 0)], the spin-splitting behavior exhibits an opposite spin polarization between spin-up and spin-down bands [**Fig. 4(d)**].

Upon further relaxing the spin structure with SOC, we find a net wFM along both the *y* and *z* axes, which is incompatible with the *Pmn*2$_1$ symmetry. This may be caused by the frustrated spin ordering along the weak spin components, due to the competition among the symmetric spin exchange interactions, Dzyaloshinskii–Moriya interactions and single-ion anisotropy. We find that the spin frustration only affects the wFM components, but not the principal spin component along the *x* direction. Therefore, although the overall magnetic symmetry reduces to *P*1 with SOC effects, the some physical phenomena we discuss below arise from the altermagnetism,of the principal spin components. Last, our Monte Carlo simulations predict a Néel temperature of ~50K (**Fig. 2(c) and Fig. S5**(*25*)).

Next, we explore possible ME coupling. Our group theory analysis reveals that the *Pmn*2$_1$ phase is a subgroup of *Cmc*2$_1$; thus, a *Pmn*2$_1$-to-*Cmc*2$_1$ phase transition may occur. The *Cmc*2$_1$ phase is close in energy to $Pmn2_1$, and our climbing nudged elastic bands calculations(*55*) also indicate that the energy barrier between the two phases is as low as 9 meV/f.u., which favors the phase transition [**Fig. 5(d)**]. The $Pmn2_1$-to-$Cmc2_1$ transition can be regarded as a ferri-to-ferroelectric transition, which can be realized by applying an electric field through its coupling to the polarizations in the two phases. We also calculated the energy barriers between $Pmn2_1$ and the other low-energy phases, such as *Pbcn*, *Pnma*, and a 90° twin domain of $Pmn2_1$. Although the $Pmn2_1$-to-*Pnma* transition also has a low energy barrier close to that of the $Pmn2_1$-to-$Cmc2_1$ transition, the energy contribution from the ***P·E*** term may favor the $Pmn2_1$-to-$Cmc2_1$ transition.

In the $Cmc2_1$ phase, the magnetic structure has A-type spin ordering. Our DFT+*U*+SOC calculations show that the $Cmc2_1$ phase has an in-plane MA perpendicular to the polarization (**Fig. S4(b)**(*25*)). Then, a magnetic group of $C_pmc2_1$ (in OG notation) is obtained [**Fig. 5(b)**]. Thus, there will be a MA change in-plane from the *x* to *y* direction between the ferrielectric and ferroelectric phases. Furthermore, there is $t\mathcal{T}$ ($\mathcal{T}$: time-reversal symmetry) in the $C_pmc2_1$ phase among the antiparallel spins; thus, the $C_pmc2_1$ phase is not altermagnetic. Since there is also spin frustration along the weak ferromagnetic spin components in the $C_pmc2_1$ phase, the overall spin structure with SOC has *P*1 symmetry [**Fig. 5(c)**].

Because both phases have *P*1 symmetries along the wFM ordering directions, the $Pmn2_1$-to-$Cmc2_1$ transition can lead to a change of both the MA directions and net wFM:~0.005 $\mu_B$/f.u., ~-0.005 $\mu_B$/f.u. and ~0.2×$10^{-4}$ $\mu_B$/f.u. along the *x*, *y*, *z* directions, respectively [**Figs. 5(e) and 5(f)**]. Furthermore, the two phases having *P*1 magnetic groups will have nonzero piezomagnetic effects, where strain may enhance the changes in the wFM at the $Pmn2_1$-to-$Cmc2_1$ transition. At -1% strain, the magnetic ordering of the principal spin components in the $Pmn2_1$ phase follows $Pmn'2_1'$ symmetry, where the MA direction is along the *y* direction. The magnetic ordering of the principal spin components in the $Cmc2_1$ phase becomes $C_pmc'2_1'$, where the MA direction is along the *x* direction (**Fig. S10** (*25*)). There is an enhancement of the wFM differences along the *z* direction between the two phases, which is 0.037 $\mu_B$/f.u. [**Figs. 5(e) and 5(f)**]. At 1% strain, the magnetic ordering of the principal spin components in the $Pmn2_1$ phase maintains $Pmn2_1$ symmetry. The magnetic ordering of the principal spin components in the $Cmc2_1$ phase is the same as that at -1% strain (**Fig. S10** (*25*)). The change of wFM becomes 0.003 $\mu_B$/f.u. along the *z* direction [**Figs. 5(e) and 5(f)**].

The enhancement of the wFM differences imposed by strain at the transition lifting the $t\mathcal{T}$ symmetry always occurs, because the piezomagnetic effects will be nonzero upon $t\mathcal{T}$ symmetry breaking and zero with $t\mathcal{T}$ symmetry. For example, if there is no spin frustration in $K_3Cr_2F_7$, strain would still enhance the wFM differences between the $Pmn'2_1'$ and $C_pmc'2_1'$ phases at -1% strain, because $t\mathcal{T}$ is broken in $Pmn'2_1'$ while it is preserved in $C_pmc'2_1'$. This explains why there are large strain induced wFM differences at -1% strain, which may be due to piezomagnetism resulting in a contribution to wFM by strain along the *z* direction (**Table S7** (*25*)). Therefore, ferroelectric control of phase transitions between altermagnetic and conventional AFM states can induce strong ME coupling in multiferroic materials, as governed by symmetry considerations. Our proposed $Pmn2_1$-to-$Cmc2_1$ transition should is expected to be generally applicable in RP materials with appropriate tolerance factors and inclusion of JT-active ions.

The piezomagnetic induced ME coupling is also observed under pressure for the $Pmn2_1$ phase. As shown in **Figs. 6(a)** and **6(b)**, a phase transition from $Pmn2_1$ to $Pbcn$ occurs at a pressure between 5 and 6 GPa. Both phases have A-type spin orderings at 5 and 6 GPa, and their magnetic symmetries are $Pmn'2_1'$ and $Pbc'n$, respectively. The altermagnetism is kept under the pressure in the $Pmn'2_1'$ phase, again due to the lack of inversion and $t\mathcal{T}$ symmetries and the existence of the spin space symmetry $[C2||\{m_{001}|t\}]$ among the antiparallel spins. In the $Pbcn$ phase, there the $t\mathcal{T}$ symmetries among the antiparallel spins prevents the existence of the altermagnetism. In our DFT+$U$+SOC calculations on both phases at 0 and 5 GPa, we find a large change of wFM with pressure in the $Pmn2_1$ phase along the *y* and *z* directions, while the changes of wFM with pressure in the $Pbcn$ phase are about 10-20 times smaller than those in the $Pmn2_1$ phase [**Figs. 6(c) and 6(d)**]. This behavior can be attributed to piezomagnetic effects present in the altermagnetic phase of $Pmn2_1$ along the *z* direction. In contrast, such effects are either absent ($Pbcn$) or significantly weaker, where the piezomagnetism arises from weak spin components that lower the symmetry to $P1$ with SOC as found in the $Pmn2_1$ and $Cmc2_1$ phases. The variation of the wFM with pressure along the *y* direction in $Pmn2_1$ can also be attributed to the weak spin components that reduce the symmetry to $P1$ upon considering SOC. The changes of the wFM at 5 GP are 0.012 $\mu_B$/f.u. Piezomagnetic effects have recently been observed in MnTe derived compounds , and a polar-to-nonpolar phases transition has also been reported in RP $Sr_3Sn_2O_7$ (*56*). These findings support the feasibility of our proposed mechanism, where ferroelectric tunablility of wFM, induced by piezomagnetism, may be experimentally realizable in future studies.

**Conclusion**

Using group theory analysis and DFT+$U$+SOC calculations, we propose a design rule for obtaining ferrielectric-to-ferroelectic phases transition in RP halides, which facilitates ferroelectric control of altermagnetic-to-conventional AFM phase transitions. At the phases transition, a strong ME coupling between the electric polarization and wFM emerges under moderate strain and pressure, driven by piezomagnetic effects in the altermagnetic phase. Our findings demonstrate the potential for realizing strong ME coupling in real space within altermagnetic compounds, paving the way for the discovery of novel functionalities by exploiting altermagnetism.

## Acknowledgments

H.-M.Z. was supported by the National Natural Science Foundation of China (Grant Nos. 12347185), the Postdoctoral Fellowship Program of CPSF under Grant Number GZC20230443，and the Jiangsu Funding Program for Excellent Postdoctoral Talent. Y.Z., H.-M.Z., C.A.J. and X.-Z.L. were supported by the National Natural Science Foundation of China (NSFC) under Grant No. 12474081, the open research fund of Key Laboratory of Quantum Materials and Devices (Southeast University), Ministry of Education, the Start-up Research Fund of Southeast University. J.M.R. was supported by the National Science Foundation (NSF) under DMR-2413680. DFT calculations were performed through computational resources and staff contributions provided for the Quest high performance computing facility at Northwestern University which is jointly supported by the Office of the Provost, the Office for Research, and Northwestern University Information Technology. Part of the calculations were performed on high-performance computers, supported by the Big Data Computing Center of Southeast University.

## References

1. L. Šmejkal, J. Sinova, T. Jungwirth, Emerging Research Landscape of Altermagnetism. *Phys. Rev. X.* **12**, 040501 (2022).

2. S. W. Cheong, F. T. Huang, Altermagnetism with non-collinear spins. *npj Quantum Mater.* **9**, 13 (2024).

3. J. Krempaský, L. Šmejkal, S. W. D'Souza, M. Hajlaoui, G. Springholz, K. Uhlířová, F. Alarab, P. C. Constantinou, V. Strocov, D. Usanov, W. R. Pudelko, R. González-Hernández, A. Birk Hellenes, Z. Jansa, H. Reichlová, Z. Šobáň, R. D. Gonzalez Betancourt, P. Wadley, J. Sinova, D. Kriegner, J. Minár, J. H. Dil, T. Jungwirth, Altermagnetic lifting of Kramers spin degeneracy. *Nature.* **626**, 517–522 (2024).

4. L. Šmejkal, J. Sinova, T. Jungwirth, Beyond Conventional Ferromagnetism and Antiferromagnetism: A Phase with Nonrelativistic Spin and Crystal Rotation Symmetry. *Phys. Rev. X.* **12**, 031042 (2022).

5. S. Lee, S. Lee, S. Jung, J. Jung, D. Kim, Y. Lee, B. Seok, J. Kim, B. G. Park, L. Šmejkal, C. J. Kang, C. Kim, Broken Kramers Degeneracy in Altermagnetic MnTe. *Phys. Rev. Lett.* **132**, 036702 (2024).

6. T. Aoyama, K. Ohgushi, Piezomagnetic properties in altermagnetic MnTe. *Phys. Rev. Mater.* **8**, L041402 (2024).

7. I. Mazin, Editorial: Altermagnetism - A New Punch Line of Fundamental Magnetism. *Phys. Rev. X.* **12**, 040002 (2022).

8. Y. P. Zhu, X. Chen, X. R. Liu, Y. Liu, P. Liu, H. Zha, G. Qu, C. Hong, J. Li, Z. Jiang, X. M. Ma, Y. J. Hao, M. Y. Zhu, W. Liu, M. Zeng, S. Jayaram, M. Lenger, J. Ding, S. Mo, K. Tanaka, M. Arita, Z. Liu, M. Ye, D. Shen, J. Wrachtrup, Y. Huang, R. H. He, S. Qiao, Q. Liu, C. Liu, Observation of plaid-like spin splitting in a noncoplanar antiferromagnet. *Nature.* **626**, 523–528

(2024).

9. Y. Guo, H. Liu, O. Janson, I. C. Fulga, J. van den Brink, J. I. Facio, Spin-split collinear antiferromagnets: A large-scale ab-initio study. *Mater. Today Phys.* **32**, 100991 (2023).

10. E. I. Rashba, Properties of semiconductors with an extremum loop .1. Cyclotron and combinational resonance in a magnetic field perpendicular to the plane of the loop. *Sov. Phys. Solid. State*. **2**, 1109 (1960).

11. G. Dresselhaus, Spin-Orbit Coupling Effects in Zinc Blende Structures. *Phys. Rev.* **100**, 580 (1955).

12. L. Šmejkal, R. González-Hernández, T. Jungwirth, J. Sinova, Crystal time-reversal symmetry breaking and spontaneous Hall effect in collinear antiferromagnets. *Sci. Adv.* **6**, eaaz8809 (2020).

13. M. Gu, Y. Liu, H. Zhu, K. Yananose, X. Chen, Y. Hu, A. Stroppa, Q. Liu, Ferroelectric switchable altermagnetism. *Phys. Rev. Lett.* **134**, 106802 (2025).

14. X. Duan, J. Zhang, Z. Zhang, I. Zutic, T. Zhou, Antiferroelectric Altermagnets: Antiferroelectricity Alters Magnets. **134**, 106801 (2025).

15. L. Šmejkal, Altermagnetic multiferroics and altermagnetoelectric effect (available at http://arxiv.org/abs/2411.19928).

16. R. Ramesh, N. A. Spaldin, Multiferroics: Progress and prospects in thin films. *Nat. Mater.* **6**, 21–29 (2007).

17. N. A. Spaldin, R. Ramesh, Advances in magnetoelectric multiferroics. *Nat. Mater.* **18**, 203–212 (2019).

18. M. Fiebig, T. Lottermoser, D. Meier, M. Trassin, The evolution of multiferroics. *Nat. Rev. Mater.* **1**, 16046 (2016).

19. N. A. Benedek, C. J. Fennie, Hybrid improper ferroelectricity: A mechanism for controllable polarization-magnetization coupling. *Phys. Rev. Lett.* **106**, 107204 (2011).

20. N. A. Benedek, M. A. Hayward, Hybrid improper ferroelectricity: A theoretical, computational and synthetic perspective. *Annu. Rev. Mater. Res.* **52**, 331 (2022).

21. J. P. Perdew, A. Ruzsinszky, G. I. Csonka, O. A. Vydrov, G. E. Scuseria, L. A. Constantin, X. Zhou, K. Burke, Restoring the density-gradient expansion for exchange in solids and surfaces. *Phys. Rev. Lett.* **100**, 136406 (2008).

22. G. Kresse, J. Furthmüller, Efficiency of ab-initio total energy calculations for metals and semiconductors using a plane-wave basis set. *Comput. Mater. Sci.* **6**, 15–50 (1996).

23. G. Kresse, J. Furthmüller, Efficient iterative schemes for *ab initio* total-energy calculations using a plane-wave basis set. *Phys. Rev. B*. **54**, 11169 (1996).

24. S. Dudarev, G. Botton, Electron-energy-loss spectra and the structural stability of nickel oxide: An LSDA+U study. *Phys. Rev. B*. **57**, 1505–1509 (1998).

25. See the Supplemental Material at http://link.aps.org/supplemental/... for more details about structural and magnetic ground states of the studied n=2 RP halides, magnetic property and electronic structure of the ferrielectric and ferroelectric phases, magnetic symmetry analysis, strain effects on the ferrielectric and ferroelectric phases, and pressure effects on the ferrielectric phase.

26. R. D. King-Smith, D. Vanderbilt, Theory of polarization of crystalline solids. *Phys. Rev. B.* **47**, 1651 (1993).

27. H. J. Xiang, E. J. Kan, S. H. Wei, M. H. Whangbo, X. G. Gong, Predicting the spin-lattice order of frustrated systems from first principles. *Phys. Rev. B.* **84**, 224429 (2011).

28. H. Xiang, C. Lee, H. J. Koo, X. Gong, M. H. Whangbo, Magnetic properties and energy-mapping analysis. *Dalt. Trans.* **42**, 823–853 (2012).

29. K. Hukushima, K. & Nemoto, Exchange Monte Carlo Method and Application to Spin Glass Simulations. *J Phys. Soc Japan.* **65**, 1604–1608 (1996).

30. F. Lou, X. Y. Li, J. Y. Ji, H. Y. Yu, J. S. Feng, X. G. Gong, H. J. Xiang, PASP: Property analysis and simulation package for materials. *J. Chem. Phys.* **154**, 114103 (2021).

31. A. Togo, I. Tanaka, First principles phonon calculations in materials science. *Scr. Mater.* **108**, 1–5 (2015).

32. O. Hellman, I. A. Abrikosov, S. I. Simak, Lattice dynamics of anharmonic solids from first principles. *Phys. Rev. B.* **84**, 180301 (2011).

33. H. Olle, I.A.Abrikosov, Temperature-dependent effective third-order interatomic force constants from first principles. *Phys.Rev.B.* **88**, 144301 (2013).

34. B. J. Campbell, H. T. Stokes, D. E. Tanner, D. M. Hatch, ISODISPLACE: a web-based tool for exploring structural distortions. *J. Appl. Crystallogr.* **39**, 607–614 (2006).

35. T. Zhu, X. Z. Lu, T. Aoyama, K. Fujita, Y. Nambu, T. Saito, H. Takatsu, T. Kawasaki, T. Terauchi, S. Kurosawa, A. Yamaji, H. B. Li, C. Tassel, K. Ohgushi, J. M. Rondinelli, H. Kageyama, Thermal multiferroics in all-inorganic quasi-two-dimensional halide perovskites. *Nat. Mater.* **23**, 182–188 (2024).

36. X.-Z. Lu, H.-M. Zhang, Y. Zhou, T. Zhu, H. Xiang, S. Dong, H. Kageyama, J. M. Rondinelli, Out-of-plane ferroelectricity and robust magnetoelectricity in quasi-two-dimensional materials. *Sci. Adv.* **9**, eadi0138 (2023).

37. A. T. Mulder, N. A. Benedek, J. M. Rondinelli, C. J. Fennie, Turning ABO3 Antiferroelectrics into Ferroelectrics: Design Rules for Practical Rotation-Driven Ferroelectricity in Double Perovskites and A3B2O7 Ruddlesden-Popper Compounds. *Adv. Funct. Mater.* **23**, 4810–4820 (2013).

38. X. Z. Lu, J. M. Rondinelli, Room Temperature Electric-Field Control of Magnetism in Layered Oxides with Cation Order. *Adv. Funct. Mater.* **27**, 1604312 (2017).

39. M. S. Senn, A. Bombardi, C. A. Murray, C. Vecchini, A. Scherillo, X. Luo, S. W. Cheong, Negative Thermal Expansion in Hybrid Improper Ferroelectric Ruddlesden-Popper Perovskites by Symmetry Trapping. *Phys. Rev. Lett.* **114**, 35701 (2015).

40. X. Q. Liu, J. W. Wu, X. X. Shi, H. J. Zhao, H. Y. Zhou, R. H. Qiu, W. Q. Zhang, X. M. Chen, Hybrid improper ferroelectricity in Ruddlesden-Popper Ca3(Ti,Mn)2O7 ceramics. *Appl. Phys. Lett.* **106**, 202903 (2015).

41. Y. S. Oh, X. Luo, F. T. Huang, Y. Wang, S. W. Cheong, Experimental demonstration of hybrid improper ferroelectricity and the presence of abundant charged walls in (Ca,Sr)3Ti2O7 crystals. *Nat. Mater.* **14**, 407–413 (2015).

42. M. J. Pitcher, P. Mandal, M. S. Dyer, J. Alaria, P. Borisov, H. Niu, J. B. Claridge, M. J.

Rosseinsky, Tilt engineering of spontaneous polarization and magnetization above 300 K in a bulk layered perovskite. *Science (80-. ).* **347**, 420–424 (2015).

43. L.-F. Huang, X.-Z. Lu, J. M. Rondinelli, Tunable Negative Thermal Expansion in Layered Perovskites from Quasi-Two-Dimensional Vibrations. *Phys. Rev. Lett.* **117**, 115901 (2016).

44. Y. Wang, F.-T. Huang, X. Luo, B. Gao, S.-W. Cheong, The First Room-Temperature Ferroelectric Sn Insulator and Its Polarization Switching Kinetics. *Adv. Mater.* **29**, 1601288 (2017).

45. X. Z. Lu, J. M. Rondinelli, Epitaxial-strain-induced polar-to-nonpolar transitions in layered oxides. *Nat. Mater.* **15**, 951–955 (2016).

46. R. Zhang, B. M. Abbett, G. Read, F. Lang, T. Lancaster, T. T. Tran, P. S. Halasyamani, S. J. Blundell, N. A. Benedek, M. A. Hayward, $La_2SrCr_2O_7$: Controlling the Tilting Distortions of $n = 2$ Ruddlesden-Popper Phases through A-Site Cation Order. *Inorg. Chem.* **55**, 8951–8960 (2016).

47. G. Gou, M. Zhao, J. Shi, J. K. Harada, J. M. Rondinelli, Anion ordered and ferroelectric ruddlesden-popper oxynitride $Ca_3Nb_2N_2O_5$ for visible-light-active photocatalysis. *Chem. Mater.* **32**, 2815–2823 (2020).

48. R. Zhang, A. S. Gibbs, W. Zhang, P. S. Halasyamani, M. A. Hayward, Structural Modification of the Cation-Ordered Ruddlesden-Popper Phase $YSr_2Mn_2O_7$ by Cation Exchange and Anion Insertion. *Inorg. Chem.* **56**, 9988–9995 (2017).

49. T. Zhu, F. Orlandi, P. Manuel, A. S. Gibbs, W. Zhang, P. S. Halasyamani, M. A. Hayward, Directed synthesis of a hybrid improper magnetoelectric multiferroic material. *Nat. Commun.* **12**, 4945 (2021).

50. J. Varignon, N. C. Bristowe, P. Ghosez, Electric Field Control of Jahn-Teller Distortions in Bulk Perovskites. *Phys. Rev. Lett.* **116**, 057602 (2016).

51. J. Varignon, N. C. Bristowe, E. Bousquet, P. Ghosez, Coupling and electrical control of structural, orbital and magnetic orders in perovskites. *Sci. Rep.* **5**, 15364 (2015).

52. A. Stroppa, P. Barone, P. Jain, J. M. Perez-Mato, S. Picozzi, Hybrid improper ferroelectricity in a multiferroic and magnetoelectric metal-organic framework. *Adv. Mater.* **25**, 2284–2290 (2013).

53. A. Cammarata, J. M. Rondinelli, Ferroelectricity from coupled cooperative Jahn-Teller distortions and octahedral rotations in ordered Ruddlesden-Popper manganates. *Phys. Rev. B.* **92**, 014102 (2015).

54. X. Z. Lu, J. M. Rondinelli, Tunable magnetic anisotropy in multiferroic oxides. *Phys. Rev. B.* **103**, 184417 (2021).

55. G. Henkelman, B. P. Uberuaga, H. Jónsson, A climbing image nudged elastic band method for finding saddle points and minimum energy paths. *J. Chem. Phys.* **113**, 9901 (2000).

56. K. A. Smith, S. P. Ramkumar, N. C. Harms, A. J. Clune, X. Xu, S. W. Cheong, Z. Liu, E. A. Nowadnick, J. L. Musfeldt, Revealing pressure-driven structural transitions in the hybrid improper ferroelectric $Sr_3Sn_2O_7$. *Phys. Rev. B.* **104**, 064106 (2021).

**Table 1**. The tolerance factor, space group, ferroelectric polarization, and band gap of the ground-state structures of the studied RP halides. The dash (-) indicates the absence of ferroelectricity in the ground-state phase.

| Materials | Tolerance factor | Ground state | Polarization ($\mu C/cm^2$) | Gap (eV) |
|---|---|---|---|---|
| $K_3Mn_2F_7$ | 0.9773 | *Cmcm* | - | 4.08 |
| $K_3Cr_2F_7$ | 0.9798 | *Pmn2$_1$* | 0.074 | 2.37 |
| $K_3Cd_2F_7$ | 0.9669 | *Cmc2$_1$* | 4.13 | 3.61 |
| $KAg_2Cr_2F_7$ | 0.9607 | *Pnma* | - | 0.17 |
| $K_3Mn_2Cl_7$ | 0.9662 | *Cmc2$_1$* | 1.95 | 4.30 |
| $KAg_2Mn_2Cl_7$ | 0.94028 | *Pc* | 4.03 | 2.43 |
| $KTl_2Mn_2Cl_7$ | 0.9701 | *P4$_2$/mnm* | - | 4.35 |
| $K_3Mg_2Cl_7$ | 0.9729 | *Cmc2$_1$* | 0.331 | 4.83 |
| $K_3Cu_2Cl_7$ | 0.9772 | *Pbcn* | - | 0.92 |
| $KAg_2Cu_2Cl_7$ | 0.9431 | *Pmn2$_1$* | 1.67 | 0.41 |

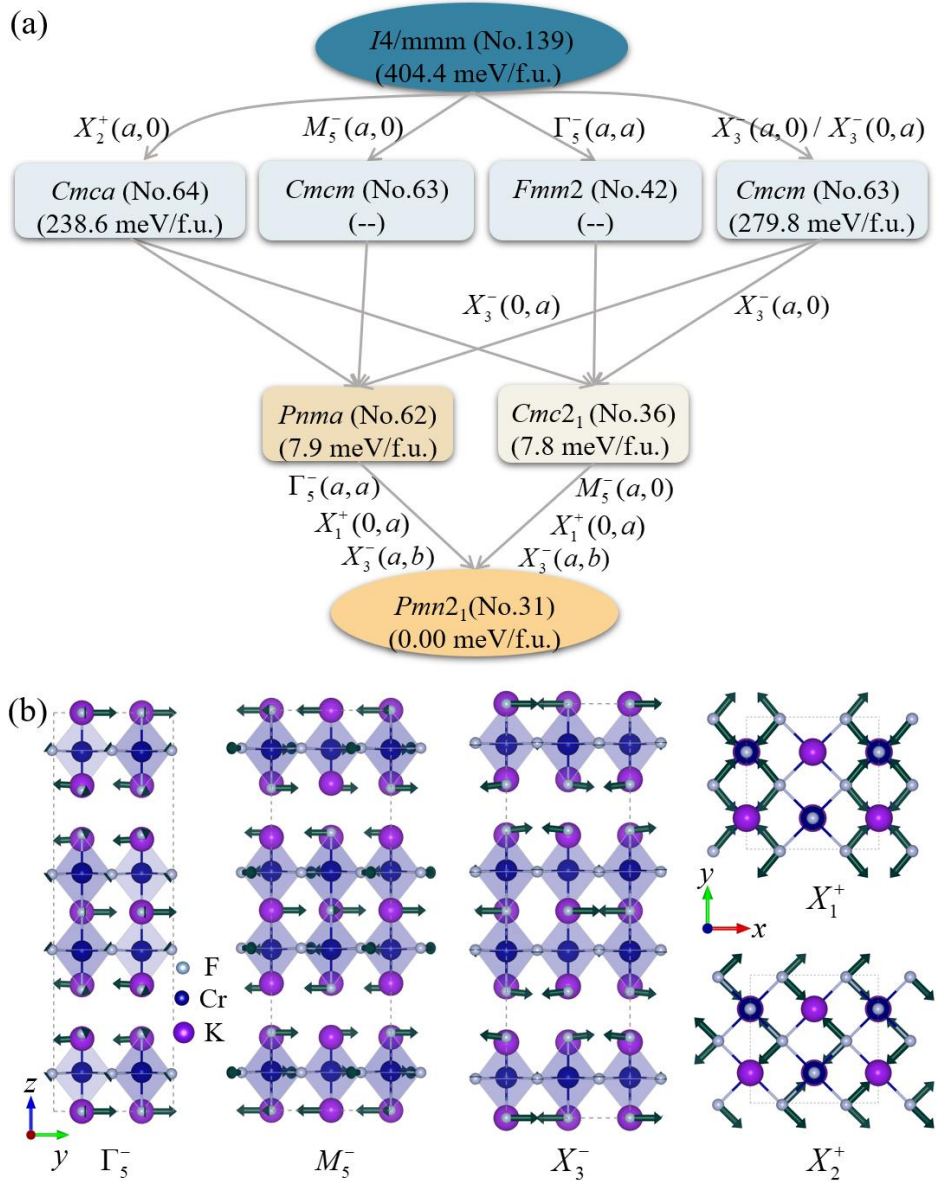

**Figure 1**. (a) Symmetries and relative energies of the structural phases by group-subgroup relationship from the high symmetry *I*4/*mmm* to *Pmn*2$_1$ phase of K$_3$Cr$_2$F$_7$. The energy of each structural phase is relative to the ground-state *Pmn*2$_1$ (No. 31) phase. The dash (--) indicates that this mode is stable during the structural optimization and the optimized structure returns to the *I*4/*mmm* phase. (b) Main atomic displacements transforming as the designated irreps of the *I*4/*mmm* (No. 139) phase of K$_3$Cr$_2$F$_7$.

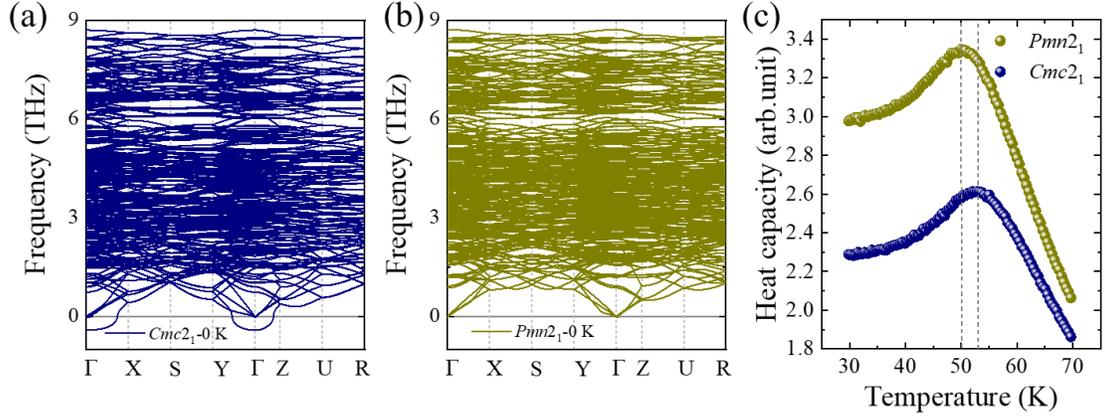

**Figure 2**. (a)-(b) Phonon spectra of $Cmc2_1$ and $Pmn2_1$ in $K_3Cr_2F_7$, respectively. There is an imaginary-frequency mode transforming as irrep $Y_3$ in the phonon dispersions of the $Cmc2_1$ phase. By adding the $Y_3$ mode into the $Cmc2_1$ phase, one obtains a stable $Pmn2_1$ phase. (c) Monte Carlo simulations on the magnetic transition temperatures of the $Cmc2_1$ and $Pmn2_1$ phases of $K_3Cr_2F_7$.

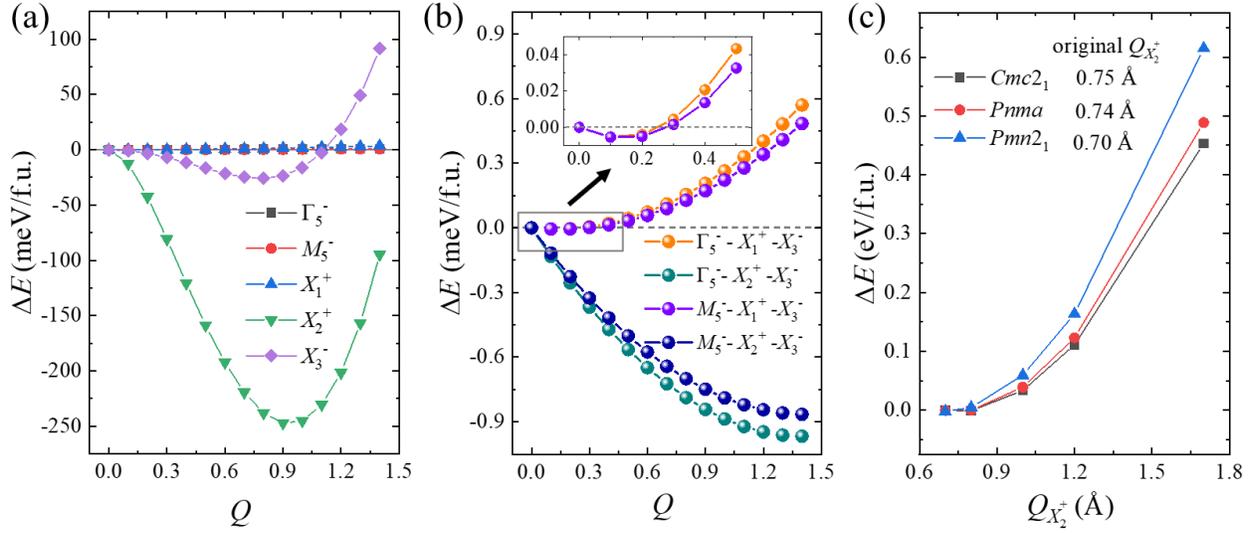

**Figure 3**. (a) Energy versus mode amplitude scaled by the original value in the equilibrium $Pmn2_1$ phase of $K_3Cr_2F_7$. (b) Energy gain arising from trilinear coupling interactions in the $Pmn2_1$ phase. The mode amplitude is scaled by the original mode amplitude in units of Å in the $Pmn2_1$ structure. (c) Energy of the $Cmc2_1$, $Pnma$ and $Pmn2_1$ phases as a function of the constrained amplitude of the $X_2^+$ mode. The original amplitudes of the $X_2^+$ mode in the three phases with fully relaxed structures are 0.75 Å, 0.74 Å, and 0.70 Å, respectively.

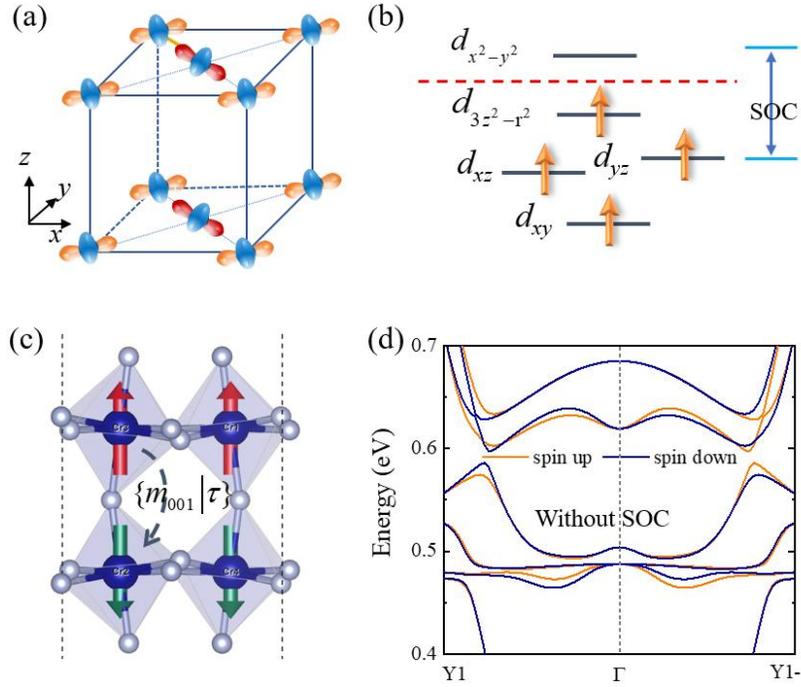

**Figure 4**. (a) Schematic illustration of the orbital ordering obtained from density of states calculations without SOC. (b) Energy levels of the $d$ orbitals of $Cr^{2+}$, where the arrows indicate orbitals occupied by spins. The main SOC interactions between the occupied and unoccupied orbitals are also shown, which leads to the easy-plane single-ion anisotropy. (c) Altermagnetic spin ordering with the antiparallel spins mapped by the spin group operation $[C_2||\{m_{001}|t\}]$. (d) Band structure of $K_3Cr_2F_7$ without SOC, the CBM is set to 0 eV. Here, there is band splitting between the spin-up and spin-down manifolds, where alternative spin-up and spin-down bands can be clearly seen in the trajectories Y1-Γ-Y1- that hosts $m_{001}$ symmetry.

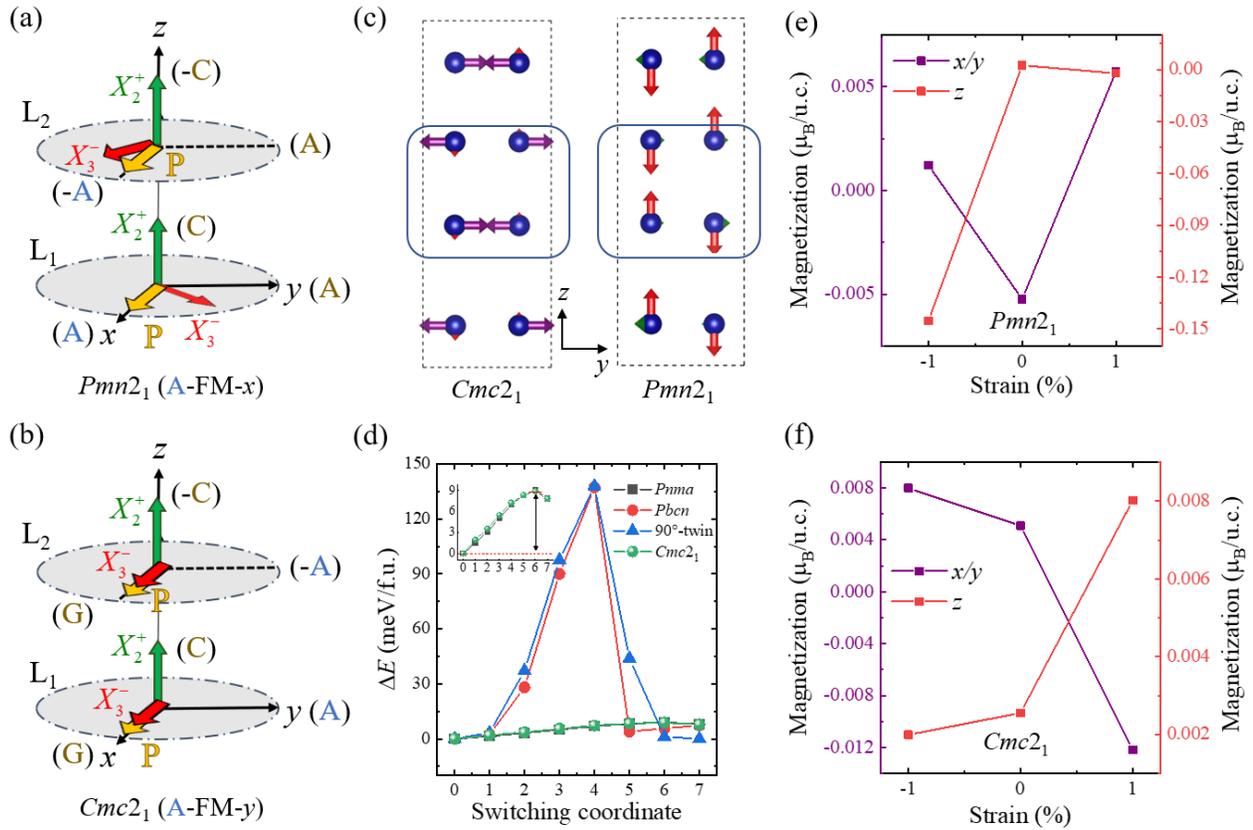

**Figure 5**. Crystal symmetry-adapted magnetic ground state structure of (a) $Pmn2_1$ and (b) $Cmc2_1$ $K_3Cr_2F_7$. $L_1$ and $L_2$ refer to the different double-perovskite layers, the blue letters refer to the principal magnetic order, the dark yellow letters represent the induced wFM orderings, and the green, red, and yellow arrows represent the octahedral rotation mode ($X_2^+$), tilt mode ($X_3^-$), and polar mode (P), respectively. "FM" and "AFM" denote the magnetic coupling between the different double-perovskite layers. (c) Calculated magnetic ground state structures with SOC for $Cmc2_1$ and $Pmn2_1$. (d) The energy barrier calculations between $Pmn2_1$ and the other low-energy phases. The changes in wFM of (e) $Pmn2_1$ and (f) $Cmc2_1$ phases under the biaxial strain. In the case of $Pmn2_1$, the wFM directions are along the $y$ and $z$ axes under 0% biaxial strain, while under tensile or compressive strain, the wFM directions are along the $x$ and $z$ axes. Conversely for $Cmc2_1$, the wFM directions are along the $x$ and $z$ axes under 0% biaxial strain, and the wFM directions are along the $y$ and $z$ axes under tensile or compressive strain. The unit cell contains eight Cr atoms.

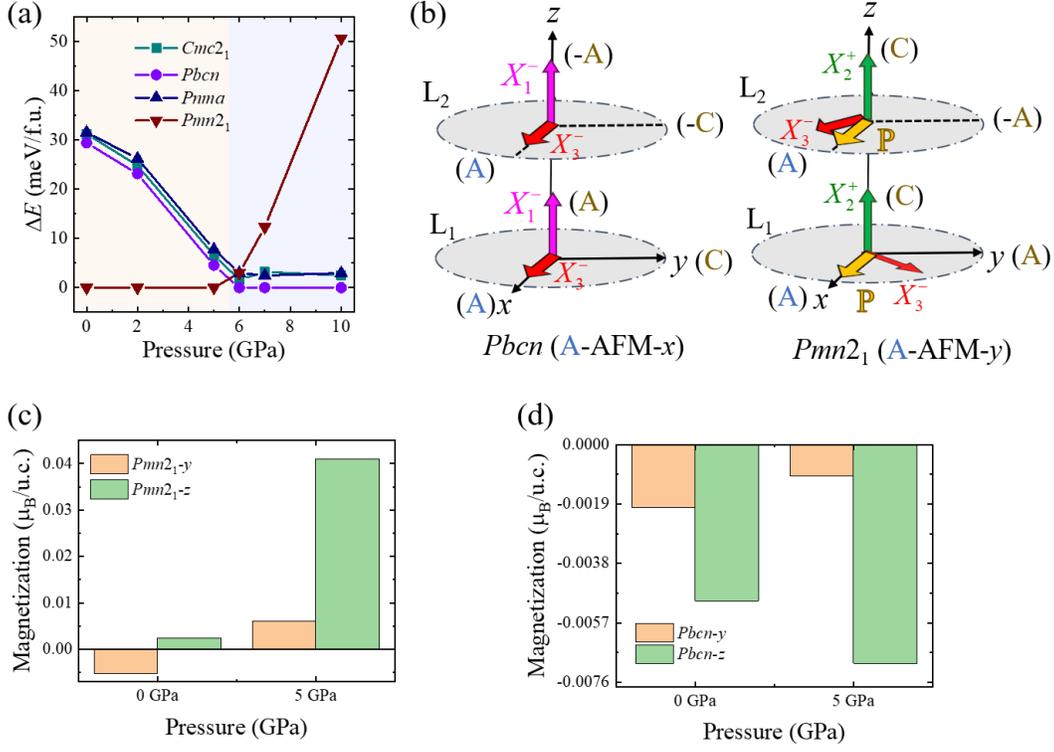

**Figure 6**. (a) Relative energies of several structural phases under pressure. The energy of $Pmn2_1$ at 0 GPa is used as a reference. At pressures less than or equal to 5 GPa, $Pmn2_1$ is the ground state, whereas at pressures greater than 5 GPa, $Pbcn$ becomes the ground state. (b) The ground state magnetic structures in $Pmn2_1$ and $Pbcn$ phases under 5 GPa. (c) and (d) show the changes in the wFM of $Pmn2_1$ and $Pbcn$ phases at 0 GPa and 5 GPa, respectively.